# GMSNP and Finite Structures

Santiago Guzmán-Pro[*][1]

[1]Institut für Algebra, TU Dresden

June 19, 2024

## Abstract

Given an (infinite) relational structure $\mathbb{S}$, we say that a finite structure $\mathbb{C}$ is a minimal finite factor of $\mathbb{S}$ if for every finite structure $\mathbb{A}$ there is a homomorphism $\mathbb{S} \to \mathbb{A}$ if and only if there is a homomorphism $\mathbb{C} \to \mathbb{A}$. In this paper we prove that if $\mathrm{CSP}(\mathbb{S})$ is in GMSNP, then $\mathbb{S}$ has a minimal finite factor $\mathbb{C}$, and moreover, $\mathrm{CSP}(\mathbb{C})$ reduces in polynomial time to $\mathrm{CSP}(\mathbb{S})$. As applications of this result, we first see that if a finite promise constraint satisfaction problem $\mathrm{PCSP}(\mathbb{A}, \mathbb{B})$ has a tractable GMSNP sandwich, then it has a tractable finite sandwich. We also show that if $\mathbb{G}$ is a non-bipartite (possibly infinite) graph with finite chromatic number, and $\mathrm{CSP}(\mathbb{G})$ is in GMSNP, then $\mathrm{CSP}(\mathbb{G})$ is NP-complete, partially answering a question recently asked by Bodirsky and Guzmán-Pro.

## 1 Introduction

Given a (possibly infinite) graph $\mathbb{G}$ we denote by $\mathrm{CSP}(\mathbb{G})$ the class of finite graphs that homomorphically map to $\mathbb{G}$. The *constraint satisfaction problem* with *template* $\mathbb{G}$ asks whether a finite input graph $\mathbb{H}$ belongs to $\mathrm{CSP}(\mathbb{G})$ — in graph theoretic terms, this is also known as the $\mathbb{G}$-colouring problem. The Hell-Nešetřil theorem [21] states that for each finite graph $\mathbb{G}$, either $\mathbb{G}$ is bipartite or has a loop (and in these cases $\mathrm{CSP}(\mathbb{G})$ is polynomial-time solvable), or $\mathrm{CSP}(\mathbb{G})$ it is NP-complete. This structural classification of the complexity of $\mathrm{CSP}(\mathbb{G})$ does not extend to the infinite case (unless P = NP): if $\mathbb{G}$ is the infinite clique, then $\mathbb{G}$ is a non-bipartite graph and $\mathrm{CSP}(\mathbb{G})$ is clearly polynomial-time solvable. It was recently noted in [9] that a (wide-open) conjecture from promise constraint satisfaction theory [14, Conjecture 1.2] implies that the Hell-Nešetřil theorem extends to infinite graphs with finite chromatic number.

**Question 1** ([9]). *Is it true that $\mathrm{CSP}(\mathbb{G})$ is NP-hard for every non-bipartite graph $\mathbb{G}$ with finite chromatic number?*

[*]santiago.guzman_pro@tu-dresden.de

The author has been funded by the European Research Council (Project POCOCOP, ERC Synergy Grant 101071674). Views and opinions expressed are however those of the author only and do not necessarily reflect those of the European Union or the European Research Council Executive Agency. Neither the European Union nor the granting authority can be held responsible for them.

In 1999, Feder and Vardi [19] conjectured that the dichotomy proved by Hell and Nešetřil for CSPs of finite graphs should generalize to constraint satisfaction problems of finite relational structures. This was confirmed independently by Zhuk [30] and Bulatov [16] in 2017. In turn, it is conjectured that this dichotomy extends to *forbidden pattern problems* [11, (particular instance of) Conjecture 1.2].

Given a finite set of edge and vertex coloured graphs $\mathcal{F}$ the forbidden pattern problem $FPP(\mathcal{F})$ is the following computation problem. On a given input finite graph $\mathbb{G}$, decide whether there is a vertex and edge colouring $\mathbb{G}'$ of $\mathbb{G}$ that (homomorphically) avoids all patterns (edge and vertex coloured graphs) in $\mathcal{F}$ — we formally introduce all these notions in the preliminary section. For instance, if $\mathcal{F}$ consist of a blue and a red edge monochromatic triangle, then $FPP(\mathcal{F})$ asks whether an input graph admits a 2-edge-colouring with no monochromatic triangles.

It is straightforward to observe that finite domain (graph) CSPs are a proper subclass of forbidden pattern problems. Actually, it suffices to consider forbidden patterns using only coloured vertices, and this subclass of FPP is captured syntactically by a logic called *monotone monadic strict NP* (MMSNP), i.e., for every set of forbidden vertex coloured patterns $\mathcal{F}$, there is an MMSNP-sentence $\phi$ such that a graph (structure) $\mathbb{G}$ satisfies $\phi$ if and only only if $\mathbb{G}$ is a yes instance of $FPP(\mathcal{F})$ [24]. So, it follows from a previous result from Feder and Vardi [18], and the finite CSP domain dichotomy [16, 30] that forbidden vertex coloured pattern problems exhibit a P vs. NP-complete dichotomy.

A natural extension of MMSNP, called $MMSNP_2$, provides a syntactic description of forbidden pattern problems with both, edge and vertex colours [3, 25]. In turn, $MMSNP_2$ is syntactically extended by the logic GMSNP, but it turns out that both logics have the same expressive power [5]; and thus, GMSNP has the same expressive power as its combinatorial counterpart described by forbidden edge and vertex coloured patterns. Contrary to the vertex colouring setting, it is still wide open whether the class of (infinite domain) CSPs expressible in GMSNP exhibits a P vs. NP-complete dichotomy. Nonetheless, it was proved in [10] that this class still falls in the scope of the *tractability conjecture* [11, Conjecture 1.2], and so it implies that if a forbidden pattern problem is not in $P$, then it is NP-complete.

In this brief note we prove a weaker form of the previous dichotomy conjecture (Corollary 14), and we show that Question 1 has a positive answer when $CSP(\mathbb{G})$ corresponds to a forbidden pattern problem. Both observations follow from our main result: for every structure $\mathbb{S}$ such that $CSP(\mathbb{S})$ is expressible in GMSNP there is a finite structure $\mathbb{C}$ with the following properties

- for every finite structure $\mathbb{A}$ there is a homomorphism $\mathbb{S} \to \mathbb{A}$ if and only if $\mathbb{C} \to \mathbb{A}$,
- $CSP(\mathbb{C})$ reduces in polynomial time to $CSP(\mathbb{S})$, and
- $\mathbb{C}$ can be constructed from any GMSNP sentence $\Phi$ defining $CSP(\mathbb{S})$.

Finally, another simple application of our main result relates to *promise constraint satisfaction problems*: if $PCSP(\mathbb{A}, \mathbb{B})$ is polynomial-time solvable by a GMSNP sandwich, then it is polynomial-time solvable by a finite CSP sandwich.

The rest of this work is structured as follows. In Section 2 we introduce most background and nomenclature needed for this work. In Section 3, we study the effect of restricting the input of CSPs to high girth instances, and we prove our main result. Finally, in Section 4 we present some applications of this result and discuss some questions that seem relevant for the scope of tractable CSP sandwiches for PCSPs [15], and for the tractability conjecture [11, Conjecture 1.2].

# 2 Preliminaries

We follow standard notions from graph theory [13]. In particular, we denote by $\mathbb{K}_n$ the complete graph on $n$ vertices. The *girth* of a graph $\mathbb{G}$ is the length of the shortest cycle in $\mathbb{G}$. We highlight that we slightly deviate from standard notation in graph theory in order to homogenize with notation for general relational structures: we use $\mathbb{G}, \mathbb{H}, \mathbb{D}, ...$ to denote graphs and digraphs, we denote by $G, H, D, ...$ the corresponding vertex sets, and by $E(\mathbb{G}), E(\mathbb{H}), E(\mathbb{D}), ...$ the respective edge sets.

## 2.1 CSPs and relational structures

A *relational signature* $\tau$ is a set of relation symbols $R, S, \ldots$ each equipped with a positive integer called its *arity*. A $\tau$-*structure* $\mathbb{A}$ consists of a *vertex set* $A$ (also called the *domain* of $\mathbb{A}$), and for each relation symbol $R \in \tau$ or arity $r$ an $r$-ary relation $R(\mathbb{A}) \subseteq A^r$ called the *interpretation* of $R$ in $\mathbb{A}$. In this context, a *digraph* $\mathbb{D}$ is an $\{E\}$-structure where $E$ is a binary relation symbol, and so, a *graph* $\mathbb{G}$ is an $\{E\}$-structure where the interpretation $E(\mathbb{G})$ is a binary symmetric relation.

Given a pair $\mathbb{A}$ and $\mathbb{B}$ of $\tau$-structures, a *homomorphism* $f \colon \mathbb{A} \to \mathbb{B}$ is a function $f \colon A \to B$ such that for every $R \in \tau$ of arity $r$ and every tuple $(a_1, \ldots, a_r) \in R(\mathbb{A})$, the tuple $(f(a_1), \ldots, f(a_r))$ belongs to the interpretation $R(\mathbb{B})$. If such a homomorphism exists, we write $\mathbb{A} \to \mathbb{B}$ and otherwise $\mathbb{A} \not\to \mathbb{B}$.

The *constraint satisfaction problem* with *template* $\mathbb{A}$ (possibly infinite) receives as input a finite structure $\mathbb{B}$ of the same signature as $\mathbb{A}$, and the task is to decide if $\mathbb{B} \to \mathbb{A}$. Following this nomenclature, we denote by $\text{CSP}(\mathbb{A})$ the class of finite structure $\mathbb{B}$ with $\mathbb{B} \to \mathbb{A}$. For instance, $\text{CSP}(\mathbb{K}_n)$ corresponds to the $n$-colourability problem (up to interpreting edges $(x, y)$ as undirected edges $xy$).

**Theorem 2** (Hell-Nešetřil theorem [21]). *Let $\mathbb{G}$ be a finite graph. If $\mathbb{G}$ is a bipartite graph or contains a loop, then $\text{CSP}(\mathbb{G})$ is polynomial-time solvable; otherwise, $\text{CSP}(\mathbb{G})$ is NP-complete*

As mentioned in the introduction, the dichotomy stated in the Hell-Nešetřil theorem generalizes to finite domain CSP. In this general setting, there is (currently) no structural characterization of the dividing line between polynomial-time tractable cases and NP-hard ones. Nonetheless, there is an algebraic and logic description of this boarder, which in particular implies that it is decidable (by a Turing machine) to test on which side of the boarder $\text{CSP}(\mathbb{A})$ lies. (Both, the algebraic and logic definitions lie outside the scope of this paper, but we include the logic description for the sake of completeness.)

**Theorem 3** ([16, 30]). *Let $\mathbb{A}$ if a finite structure with a finite relational signature $\tau$. If $\text{CSP}(\mathbb{A})$ is not polynomial-time solvable, then $\mathbb{A}$ primitively positively constructs $K_3$, and in this case $\text{CSP}(\mathbb{A})$ is NP-complete.*

## 2.2 Duality pairs

Given a set of (possibly infinite) structures $\mathcal{F}$, we denote by $\text{Forb}(\mathcal{F})$ the class of finite structures $\mathbb{A}$ such that $\mathbb{F} \not\to \mathbb{A}$ for every $\mathbb{F} \in \mathcal{F}$. When $\mathcal{F} = \{\mathbb{B}\}$, we simply write $\text{Forb}(\mathbb{B})$. A *finite duality pair* $(\mathcal{F}, \mathbb{D})$ consists of a finite set $\mathcal{F}$ and a structure $\mathbb{D}$ such that $\text{Forb}(\mathcal{F}) = \text{CSP}(\mathbb{D})$. A well-known family of examples comprise *transitive tournaments* on $n$ vertices $\mathbb{T}_n$, i.e., the digraph with vertex set $\{1, \ldots, n\}$ and $(i, j) \in E(\mathbb{T}_n)$ if and only if $i < j$, and $\overrightarrow{\mathbb{P}}_n$ the directed path on $n$ vertices. For every positive integer $n$, the pair $(\{\overrightarrow{\mathbb{P}}_{n+1}\}, \mathbb{T}_n)$ is a finite duality pair [6, Theorem 3.1].

Given a structure $\mathbb{A}$ with finite relational signature $\tau$, the *incidence graph* of $\mathbb{A}$ is the bipartite graph $(A, \cup_{R\in\tau} R(\mathbb{A}))$ and there is an edge $(a, (a_1, \ldots, a_r))$ if $a = a_i$. Notice that this generalizes the standard notion of the incidence graph of a graph $\mathbb{G}$. We say that $\mathbb{A}$ is a *tree* if its incidence graph is a tree. Similarly, we say that $\mathbb{A}$ is *connected* if its incidence graph if connected, and the *girth* of $\mathbb{A}$ is defined as half the girth of its incidence graph. Trees and finite duality pairs are closely related as the following statement asserts.

**Theorem 4** ([27]). *For ever finite set of finite trees $\mathcal{T}$ with finite signature $\tau$, there is a finite $\tau$-structure $\mathbb{D}$ such that $\mathrm{Forb}(\mathcal{T}) = \mathrm{CSP}(\mathbb{D})$.*

## 2.3 Forbidden pattern problems

Consider a finite relational signature $\tau$, a finite set of vertex colours $\mathcal{V}$, and for each $R \in \tau$ a finite set of $R$-colours $\mathcal{R}$. A $(\mathcal{V}, \{\mathcal{R}\}_{R\in\tau})$-*colouring* of a $\tau$-structure $\mathbb{A}$ is function $c_V \colon A \to \mathcal{V}$ together with a function $c_R \colon R(\mathbb{A}) \to \mathcal{R}$ for each $R \in \tau$. When there is no risk of ambiguity, we will simply talk about a *colouring* of $\mathbb{A}$ (instead of a $(\mathcal{V}, \{\mathcal{R}\}_{R\in\tau})$-colouring).

Notice that each colouring of a $\tau$-structure $\mathbb{A}$ can be regarded as a structure with signature $\cup_{R\in\tau}\mathcal{R} \cup \mathcal{V}$ where each $V_i \in \mathcal{V}$ is a unary relational symbol, and each $R_i \in \mathcal{R}$ has the same arity as $R$. So, the interpretation of each colour $R_i \in \mathcal{R}$ is $c_R^{-1}(R_i)$ and of each unary symbol $V_i \in \mathcal{V}$ is $c_V^{-1}(V_i)$. Actually, for the present work, a colouring of a $\tau$-structure will be identified with the relational structure $\mathbb{A}'$ previously defined, and we say that $\mathbb{A}'$ is a *coloured $\tau$-structure*. In particular, when we talk about *colour-preserving homomorphism* between a colouring of $\mathbb{A}$ and a colouring of $\mathbb{B}$, we simply refer to a homomorphism $\mathbb{A}' \to \mathbb{B}'$ of the corresponding relational structures.

A $\tau$-*pattern* is a colouring $\mathbb{F}'$ of a connected finite $\tau$-structure $\mathbb{F}'$. Given a finite set of patterns $\mathcal{F}$ (with colours $(\mathcal{V}, \{\mathcal{R}\}_{R\in\tau})$), the *forbidden pattern problem* takes as an input a finite $\tau$-structure $\mathbb{A}$ and the task is to decide if there is a $(\mathcal{V}, \{\mathcal{R}\}_{R\in\tau})$-colouring $\mathbb{A}'$ of $\mathbb{A}$ such that $\mathbb{A}' \in \mathrm{Forb}(\mathcal{F})$, i.e., for every $\mathbb{F}' \in \mathcal{F}$, there is no (colour-preserving) homomorphism $\mathbb{F}' \not\to \mathbb{A}'$. Similarly as we did we CSPs, We will by $\mathrm{FPP}(\mathcal{F})$ the class of yes-instance to the forbidden patter problem (with forbidden pattern $\mathcal{F}$). Also, as we do with CSPs, we talk about $\mathrm{FPP}(\mathcal{F})$ being polynomial-time solvable or NP-complete the corresponding forbidden pattern problem is polynomial-time solvable or NP-complete, and we also say that $\mathrm{FPP}(\mathcal{F})$ (as a language of $\tau$-structures) is in P or in NP.

Well-known examples of forbidden pattern problems include 3-colourability (and any finite CSP), colouring the vertices of a graph with two colours in such a way that there are no monochromatic triangles, and similarly, colouring the edges of graph (with two colours) in such a way that there are no monochromatic triangles. Less obvious examples include certain forbidden orientation and orientation completion problems [2, 8, 20].

A class of finite $\tau$-structure $\mathcal{C}$ is closed or preserved under *inverse homomorphisms* if for every $\mathbb{A} \in \mathcal{C}$ and $\mathbb{B} \to \mathbb{A}$ it is the case that $\mathbb{B} \in \mathcal{C}$, and it is closed under disjoint unions if $\mathbb{A} + \mathbb{B} \in \mathcal{C}$ whenever $\mathbb{A}, \mathbb{B} \in \mathcal{C}$. It is not hard to see that a class $\mathcal{C}$ is the CSP of a (possibly infinite) structure $\mathbb{S}$ if and only if $\mathcal{C}$ is preserved under inverse homomorphisms and disjoint unions (see, e.g., [7, Lemma 1.1.8]). It is straightforward to observe that for every finite set of $\tau$-patterns $\mathrm{FPP}(\mathcal{F})$ is closed under inverse homomorphisms, and since patterns are colouring of connected structures, $\mathrm{FPP}(\mathcal{F})$ is also preserved under disjoint unions. Hence, for every finite set of $\tau$-patterns $\mathcal{F}$, there is a $\tau$-structure $\mathbb{S}$ such that $\mathrm{FPP}(\mathcal{F}) = \mathrm{CSP}(\mathbb{S})$.

## 2.4 GMSNP

We assume familiarity with first-order logic. *Guarded monotone strict NP* (GMSNP) is the following fragment of existential second order logic. Given a finite relational signature $\tau$, a $\tau$-sentence of GMNSP is of the form $\exists R_1, \ldots, R_k \forall x_1, \ldots, x_n \phi(x_1, \ldots, x_n)$ where $R_1, \ldots, R_k$ are relation symbols not in $\tau$, and $\phi$ is a conjunction $\neg \phi_1 \wedge \cdots \wedge \neg \phi_m$ of negated formulas $\neg \phi_i := \neg(\alpha_i \wedge \beta_i)$ such that:

- each $\alpha_i$ is a conjunction of positive atomic $(\{R_1, \ldots, R_k\} \cup \tau)$-formulas,
- each $\beta_i$ is a conjunction of negated atomic $\{R_1, \ldots, R_k\}$-formulas, and
- each atom $R_j(\overline{x})$ of $\beta_i$ is *guarded* by some atom $S(\overline{y})$ of $\alpha_i$, i.e., all variables in $\overline{x}$ appear in $\overline{y}$.

GMSNP syntactically generalizes a logic denoted by MMSNP$_2$ (see, e.g., [3, 25]), and it was proved in [5] that for every sentence $\Phi$ of GMSNP there is a sentence $\Psi$ of MMSNP$_2$ such that a $\tau$-structure $\mathbb{A}$ satisfies $\Phi$ if and only if it satisfies $\Psi$. In turn, it was proved in [25] that for every finite set of $\tau$-patterns $\mathcal{F}$ there is a sentence $\Psi$ of MMSNP$_2$ such that a $\tau$-structure $\mathbb{A}$ belongs to FPP($\mathcal{F}$) if and only if $\mathbb{A} \models \Psi$. Moreover, for every sentence $\Psi$ of MMSNP$_2$, there are sets of $\tau$-patterns $\mathcal{F}_1, \ldots, \mathcal{F}_m$ such that a $\tau$-structure $\mathbb{A}$ satisfies $\Psi$ if and only if $\mathbb{A}$ belongs to the union $\bigcup_{i=1}^m \text{FPP}(\mathcal{F}_i)$ [25, Corollary 9].

Given a $\tau$-structure $\mathbb{S}$, we say that CSP($\mathbb{S}$) is *expressible in* GMSNP, or simply that CSP($\mathbb{S}$) is *in* GMSNP, if there is a $\tau$-sentence $\Phi$ of GMSNP such that a finite $\tau$-structure $\mathbb{A}$ satisfies $\Phi$ if and only if $\mathbb{A} \in \text{CSP}(\mathbb{S})$. The next statement follows from the discussion in the previous paragraph.

**Theorem 5** ([5, 25]). *Let $\tau$ be a finite relational signature and $\mathbb{S}$ a $\tau$-structure. Then, CSP($\mathbb{S}$) is expressible in GMSNP if and only if there is some finite set $\mathcal{F}$ of $\tau$-patterns such that CSP($\mathbb{S}$) = FPP($\mathcal{F}$).*

Finally, it was proved in [10] that every CSP expressible in MMSNP$_2$ (equivalently, in GMSNP) is the CSP of a *reduct of a finitely bounded homogeneous* structure — this definition is not needed for the present work — and thus, such a CSP lies in the scope of the tractability conjecture (which conjecture a generalization of Theorem 3).

**Conjecture 1** (particular instance of Conjecture 1.2 in [11]). *Let $\mathbb{S}$ be a structure with a finite relational signature $\tau$ such that CSP($\mathbb{S}$) is in GMSNP. If CSP($\mathbb{S}$) is not polynomial-time solvable, then $\mathbb{S}$ primitively positive constructs $\mathbb{K}_3$, and in this case CSP($\mathbb{S}$) is NP-complete.*

## 3 Large girth

A celebrated result from Erdős [17] states that for every pair of positive integer $l, k$ there is a graph $\mathbb{G}$ with girth strictly larger than $l$ and such that $\mathbb{G}$ does not admit a proper $k$-colouring, in other words, $\mathbb{G} \not\to \mathbb{K}_k$. Actually, for every positive integer $l$ and every graph $\mathbb{H}$ there is a graph $\mathbb{G}$ of girth strictly larger than $l$ such that $\mathbb{G} \not\to \mathbb{H}$ [21, Corollary 3.14]. This result generalizes to arbitrary relational structures.

**Theorem 6** (Sparse Incomparability Lemma [23]). *Let $k$ and $l$ be positive integers and $\tau$ a finite relational signature. For every finite $\tau$-structure $\mathbb{A}$ there is a $\tau$-structure $\mathbb{B}$ with the following properties:*

- $\mathbb{B} \to \mathbb{A}$,

- *the girth of $\mathbb{B}$ is larger than $l$,*
- $\mathbb{A} \to \mathbb{C}$ *if and only if* $\mathbb{B} \to \mathbb{C}$ *for every structure $\mathbb{C}$ on at most $k$ vertices,*
- $\mathbb{B}$ *can be constructed in polynomial time (from $\mathbb{A}$).*

Given a positive integer $l$ and a structure $\mathbb{A}$, we denote by $\mathrm{CSP}_{>l}(\mathbb{A})$ the intersection of $\mathrm{CSP}(\mathbb{A})$ with structures of girth strictly larger than $l$.

**Corollary 7.** *For ever finite $\tau$-structure $\mathbb{A}$ and every positive integer $l$, $\mathrm{CSP}(\mathbb{A})$ and $\mathrm{CSP}_{>l}(\mathbb{A})$ are polynomial-time equivalent.*

A natural question that arises is whether the polynomial-time equivalence between $\mathrm{CSP}(\mathbb{A})$ and $\mathrm{CSP}_{>l}(\mathbb{A})$ extends to (some well-behaved class of) infinite structures. Unfortunately, this is not the case, not even in tame classes of infinite domain CSPs such as MMSNP: let $\mathbb{H}$ be an infinite graph such that $\mathrm{CSP}(\mathbb{H})$ is the class of graphs that admit a 2-vertex colouring without monochromatic triangles; clearly, $\mathrm{CSP}_{>3}(\mathbb{H})$ is trivial while $\mathrm{CSP}(\mathbb{H})$ is NP-complete (for instance, one can reduce from not-all-equal 3-SAT).

### 3.1 Finite-domain up to high girth

We say that the CSP of a structure $\mathbb{S}$ is *finite-domain up to high girth* if there is a finite structure $\mathbb{C}$ and a positive integer $l$ such that $\mathrm{CSP}_{>l}(\mathbb{S}) = \mathrm{CSP}_{>l}(\mathbb{C})$. In this case we say that $\mathbb{C}$ is a *finite representative* of high girth instances of $\mathrm{CSP}(\mathbb{S})$.

**Remark 8.** *If $\mathbb{C}$ is a finite representative of high girth instances of $\mathrm{CSP}(\mathbb{S})$, then $\mathrm{CSP}(\mathbb{C})$ reduces in polynomial time to $\mathrm{CSP}(\mathbb{S})$ — such a reduction can be obtained via the Sparse Incomparability Lemma (Theorem 6).*

We say that a finite structure $\mathbb{C}$ is a *minimal finite factor* of $\mathbb{S}$ if for every finite structure $\mathbb{B}$ there is a homomorphism $\mathbb{S} \to \mathbb{B}$ if and only if there is a homomorphism $\mathbb{C} \to \mathbb{B}$. It is not hard to notice that if $\mathbb{C}_1$ and $\mathbb{C}_2$ are minimal finite factors of $\mathbb{S}$, then $\mathbb{C}_1$ and $\mathbb{C}_2$ are homomorphically equivalent. A simple example of a structure that does not have a minimal finite factor is the infinite directed path $P_\omega$. Indeed, $P_\omega$ homomorphically maps to every directed cycle, but to no directed path, and clearly there is no finite digraph that homomorphically maps to all directed cycles but to no directed path. It is sensible to ask whether some tame (model theoretic) property of infinite structures $\mathbb{S}$ implies the existence of a minimal finite factor $\mathbb{C}$. For instance, this question has been considered for $\omega$-categorical structures [26]. Here, we observe that if $\mathrm{CSP}(\mathbb{S})$ is finite-domain up to high girth, then $\mathbb{S}$ has a minimal finite factor.

**Theorem 9.** *Consider a pair of structures $\mathbb{C}$ and $\mathbb{S}$. If $\mathbb{C}$ is a finite representative of high girth instances of $\mathrm{CSP}(\mathbb{S})$, then $\mathbb{C}$ is a minimal finite factor of $\mathbb{S}$.*

*Proof.* We first show that $\mathbb{S} \to \mathbb{C}$, and anticipating a contradiction, suppose that $\mathbb{S} \not\to \mathbb{C}$. So, by compactness there is some finite substructure $\mathbb{A}$ of $\mathbb{S}$ that does not map to $\mathbb{C}$. Then, by the Sparse Incomparability Lemma, for every positive integer $l$, there is a structure $\mathbb{B}$ of girth larger than $l$ such that $\mathbb{B} \to \mathbb{A}$ and $\mathbb{A} \not\to \mathbb{C}$. The latter implies that $\mathbb{B} \in \mathrm{CSP}_{>l}(\mathbb{S}) \setminus \mathrm{CSP}_{>l}(\mathbb{C})$, contradicting the choice of $\mathbb{C}$. Hence, $\mathbb{S} \to \mathbb{C}$ and thus, $\mathbb{C} \to \mathbb{A}$ implies that $\mathbb{S} \to \mathbb{A}$ for every finite structure $\mathbb{A}$.

Now, we show that if $\mathbb{C} \not\to \mathbb{A}$ for some finite structure $\mathbb{A}$, then $\mathbb{S} \not\to \mathbb{A}$. Again, by the Sparse Incomparability Lemma, for every positive integer $l$ there is a structure $\mathbb{B}_l$ with girth larger than $l$ such that $\mathbb{B}_l \to \mathbb{C}$ and $\mathbb{B}_l \not\to \mathbb{A}$. Since $\mathrm{CSP}_{>l}(\mathbb{C}) = \mathrm{CSP}_{>l}(\mathbb{S})$ for some positive integer $l$, it follows that $\mathbb{B}_{l+1} \to \mathbb{S}$, and thus $\mathbb{S} \not\to \mathbb{A}$. □

**Corollary 10.** *If* CSP($\mathbb{S}$) *is finite-domain up to high girth, then $\mathbb{S}$ has a minimal finite factor $\mathbb{C}$. In particular, any finite representative $\mathbb{C}'$ of high girth instances of* CSP($\mathbb{S}$) *is homomorphically equivalent to $\mathbb{C}$.*

It would be too surprising if the converse of (the first statement of) Corollary 10 was also true. For the sake of completion, we provide a concrete example of an ($\omega$-categorical) structure (actually, a reduct of a finitely bounded homogeneous structure) that has a minimal finite factor but its CSP is not finite-domain up to high girth.

**Example 11.** *The* generic circular triangle-free graph $\mathbb{C}_3$ *introduced in [9] has as vertex set a dense subset of the unit circle, and a pair of vertices $x, y$ are adjacent if and only if the length of each circular arc with endpoints $x$ and $y$ is strictly larger than $1/3$. It is straightforward to observe that $\mathbb{C}_3 \to K_3$ (simply partition the circle into three circular arcs of length $1/3$). It is also not hard to notice that every pair of non-adjacent vertices have a common neighbour and so, if $\mathbb{C}_3 \to G$ for some finite graph $G$, then $G$ must have a triangle. Therefore, $K_3$ is a minimal finite factor of $\mathbb{C}_3$. The generic circular triangle-free graph was introduced so $G \in \text{CSP}(\mathbb{C}_3)$ if and only if its circular chromatic number $\chi_c(G)$ is strictly less than 3 (see, e.g., [9, Corollary 13]). It was proved in [28] that for every positive integer $l$, there is a graph $G$ of girth strictly larger than $l$ and $\chi_c(G) = 3$ — and thus $G \to K_3$ (see, e.g., [29, Theorem 1.1]). Therefore, $G \in \text{CSP}_{>l}(K_3) \setminus \text{CSP}_{>l}(\mathbb{C}_3)$ and so, $\mathbb{C}_3$ has a minimal finite factor but* CSP($\mathbb{C}_3$) *is not finite-domain up to high girth. It was proved in [9] that $\mathbb{C}_3$ is an $\omega$-categorical structure, and moreover a reduct of a finitely bounded homogeneous structure.*

### 3.2 GMSNP is finite-domain up to high girth

Now we show that every CSP expressible in GMSNP is finite-domain up to high girth.

**Lemma 12.** *Let $\mathbb{S}$ be a relational structure. If* CSP($\mathbb{S}$) *is in* GMSNP*, then* CSP($\mathbb{S}$) *is finite-domain up to high girth.*

*Proof.* For this proof, we will consider the equivalent forbidden pattern problem to CSP($\mathbb{S}$) (Theorem 5). Let $\mathcal{F}$ be a finite set of $\tau$-patterns such that FPP($\mathcal{F}$) = CSP($\mathbb{S}$), and let $\sigma$ be the signature of such $\tau$-patterns. Throughout the proof, we will write $\mathbb{A}_\sigma$ for a $\sigma$-structure corresponding to a colouring of a $\tau$-structure $\mathbb{A}$.

Let $\mathcal{T}$ be the set of trees $\mathbb{T}_\sigma$ such that there is a surjective homomorphism $f \colon \mathbb{F}_\sigma \to \mathbb{T}_\sigma$ for some $\mathbb{F}_\sigma \in \mathcal{F}$. Let $\mathbb{D}_\sigma$ be a dual of $\mathcal{T}$ (Theorem 4). We first note that every structure $\mathbb{C} \in \text{CSP}(\mathbb{S})$ homomorphically maps to $\mathbb{D}$: let $\mathbb{C}_\sigma$ be a $\sigma$-colouring of $\mathbb{C}$ such that $\mathbb{C}_\sigma \in \text{Forb}(\mathcal{F})$; then $\mathbb{C}_\sigma \in \text{Forb}(\mathcal{T})$ and $\mathbb{C}_\sigma \to \mathbb{D}_\sigma$, thus $\mathbb{C} \to \mathbb{D}$. In particular, $\text{CSP}_{>l}(\mathbb{S}) \subseteq \text{CSP}_{>l}(\mathbb{D})$ for every positive integer $l$. Now, let $l$ be a positive integer larger than the number of vertices in every structure in $\mathcal{F}$. Suppose that $\mathbb{C}$ is a finite structure of girth larger than $l$ and there is a homomorphism $f \colon \mathbb{C} \to \mathbb{D}$. Notice that $f$ and the $\sigma$-colouring $\mathbb{D}_\sigma$ of $\mathbb{D}$ define a (unique) $\sigma$-colouring $\mathbb{C}_\sigma$ of $\mathbb{C}$ such that $f \colon \mathbb{C}_\sigma \to \mathbb{D}_\sigma$ is a homomorphism. We claim that $\mathbb{C}_\sigma \in \text{Forb}(\mathcal{F})$. On the contrary, suppose that there is some $\mathbb{F}_\sigma \in \mathcal{F}$ such that $\mathbb{F}_\sigma \to \mathbb{C}_\sigma$. Since the girth of $\mathbb{C}$ (and of $\mathbb{C}_\sigma$) is larger than the number of vertices of $\mathbb{F}_\sigma$, then the image of any homomorphism $h \colon \mathbb{F}_\sigma \to \mathbb{C}_\sigma$ is a tree. Thus, there is some tree in $\mathcal{T}$ homomorphically mapping to $\mathbb{C}_\sigma$, i.e., $\mathbb{C}_\sigma \notin \text{Forb}(\mathcal{T})$. This contradicts the fact that $\text{Forb}(\mathcal{T}) = \text{CSP}(\mathbb{D}_\sigma)$ and the fact that $f \colon \mathbb{C}_\sigma \to \mathbb{D}_\sigma$ is a homomorphism. Therefore, $\mathbb{C} \in \text{CSP}(\mathbb{S})$, and this concludes the proof. □

**Theorem 13.** *For every structure $\mathbb{S}$ with finite relational signature $\tau$ such that $\mathrm{CSP}(\mathbb{S})$ is in GMSNP, there is a finite structure $\mathbb{C}$ such that the following statements hold,*

- *$\mathbb{C}$ is a minimal finite factor of $\mathbb{S}$,*
- *$\mathrm{CSP}(\mathbb{C})$ reduces in polynomial time to $\mathrm{CSP}(\mathbb{S})$, and*
- *$\mathbb{C}$ can be computed from any sentence $\Phi$ in GMSNP or any set of forbidden patterns $\mathcal{F}$ such that $\mathrm{FPP}(\mathcal{F}) = \mathrm{CSP}(\mathbb{S})$.*

*Proof.* By Lemma 12, $\mathrm{CSP}(\mathbb{S})$ has a finite representative of large girth instances $\mathbb{C}$. In turn, by Corollary 10 we know that $\mathbb{C}$ is the unique minimal finite factor $\mathbb{C}$ of $\mathbb{S}$ (up to homomorphic equivalence). The polynomial-time reduction from the second statement follows from Remark 8. Finally, to see the last statement, if $\Phi$ is a sentence of GMSNP defining $\mathrm{CSP}(\mathbb{S})$, it follows from the proofs of Theorem 5 (in [5,25]) that one can compute a set $\mathcal{F}$ of $\tau$-patterns with $\mathrm{FPP}(\mathcal{F}) = \mathrm{CSP}(\mathbb{S})$. Now, from the set of patterns $\mathcal{F}$ one can compute the minimal finite factor $\mathbb{C}$ due to the constructions from the proof of Lemma 12, and from the proof of Theorem 4. □

The Sparse Incomparability Lemma asserts that the finite domain CSP dichotomy is equivalent to the following statement: for every finite structure $\mathbb{A}$, if there is no positive integer $l$ such that $\mathrm{CSP}_{>l}(\mathbb{A})$ is in P, then $\mathrm{CSP}_{>l}(\mathbb{A})$ is NP-complete for every positive integer $l$. It follows from our results that the previous dichotomy does extend to GMSNP. Notice that as observed above, the Sparse Incomparability Lemma does not extend to the infinite setting (not even to MMSNP), so this does not prove the dichotomy conjecture for GMSNP.

**Corollary 14.** *Let $\mathbb{S}$ be a structure with a finite signature $\tau$ such that $\mathrm{CSP}(\mathbb{S})$ is in GMSNP. If there is no positive integer $l$ such that $\mathrm{CSP}_{>l}(\mathbb{S})$ is in P, then $\mathrm{CSP}_{>l}(\mathbb{S})$ is NP-complete for every positive integer $l$. Moreover, it is decidable to which class $\mathrm{CSP}(\mathbb{S})$ belongs given any sentence $\Phi$ in GMSNP defining $\mathrm{CSP}(\mathbb{S})$.*

*Proof.* By Theorem 13 there is a minimal finite factor $\mathbb{C}$ of $\mathbb{S}$. From Lemma 12 and Corollary 10, there is a positive integer $l$ such that $\mathrm{CSP}_{>l}(\mathbb{C}) = \mathrm{CSP}_{>l}(\mathbb{S})$. Thus, the first statement follows from the Sparse Incomparability Lemma. The last one holds because $\mathbb{C}$ is computable from $\Phi$, and thus, the finite CSP dichotomy theorem 3 guarantees that we can decide whether $\mathrm{CSP}(\mathbb{C})$ is polynomial-time solvable or NP-complete. □

## 4 Applications and discussion

### 4.1 PCSPs

Given a pair of $\tau$-structures $\mathbb{A}$ and $\mathbb{B}$ with $\mathbb{A} \to \mathbb{B}$, the *promise constraint satisfaction problem* $\mathrm{PCSP}(\mathbb{A}, \mathbb{B})$ is the following computational problem. The input space consists of finite $\tau$-structures $\mathbb{C}$, and the task is to distinguish the cases where $\mathbb{C} \to \mathbb{A}$ and when $\mathbb{C} \not\to \mathbb{B}$. In other words, output 'yes' if $\mathbb{C} \to \mathbb{A}$, output 'no' if $\mathbb{C} \not\to \mathbb{B}$, and the answer can be arbitrary whenever $\mathbb{C} \not\to \mathbb{A}$ and $\mathbb{C} \to \mathbb{B}$. This class of problems extends finite domain CSPs since $\mathrm{PCSP}(\mathbb{A}, \mathbb{A})$ is the same problem as $\mathrm{CSP}(\mathbb{A})$.

A standard technique, sometimes called *sandwich technique* for solving $\mathrm{PCSP}(\mathbb{A}, \mathbb{B})$ in polynomial time is to find a (possibly infinite) structure $\mathbb{S}$ such that $\mathbb{A} \to \mathbb{S} \to \mathbb{B}$ and $\mathrm{CSP}(\mathbb{S})$ is in P.

Moreover, it was conjectured in [15] that this technique is necessary and sufficient for polynomial-time tractability of finite domain PCSPs.

It has been proved that in some cases, the sandwich technique needs and infinite domain CSP (unless P = NP). For instance, it is known that PCSP(1-IN-3, NAE)[1] is solvable in polynomial time via the sandwich 1-IN-3 $\to (\mathbb{Z}, \{x + y + z = 1\}) \to$ NAE, and it was proved in [4] that CSP($\mathbb{S}$) is NP-complete for every finite structure $\mathbb{S}$ such that 1-IN-3 $\to \mathbb{S} \to$ NAE. It it open whether in this case, the integers can be substituted by an $\omega$-categorical structure [4].

A nice application of Theorem 13 is that GMSNP is only as powerful as finite domain CSP for the sandwich technique, i.e., if there is a tractable GMSNP sandwich for PCSP($\mathbb{A}, \mathbb{B}$), then there is a tractable finite sandwich for PCSP($\mathbb{A}, \mathbb{B}$).

**Theorem 15.** *The following statements are equivalent for every pair $\mathbb{A}$, $\mathbb{B}$ of finite structures with finite relational signature*

- *there is a finite sandwich $\mathbb{A} \to \mathbb{C} \to \mathbb{B}$ such that CSP($\mathbb{C}$) is in P, and*
- *there is a sandwich $\mathbb{A} \to \mathbb{C} \to \mathbb{B}$ such that CSP($\mathbb{C}$) is in P and in GMSNP.*

*Proof.* The first item clearly implies the second one. Conversely, let $\mathbb{S}$ be such a structure, and $\mathbb{C}$ be a minimal finite factor of $\mathbb{S}$ (Theorem 13). Then, we know that $\mathbb{A} \to \mathbb{S} \to \mathbb{C} \to \mathbb{B}$, and CSP($\mathbb{C}$) reduces in polynomial time to CSP($\mathbb{S}$). The claim now follows. □

**Corollary 16.** *Let $\mathbb{S}$ be an infinite structure such that 1-IN-3 $\to \mathbb{S} \to$ NAE. If CSP($\mathbb{S}$) is expressible in GMSNP, then CSP($\mathbb{S}$) is NP-complete.*

*Proof.* Direct application from Theorem 15 and the fact that every finite sandwich 1-IN-3 $\to \mathbb{S} \to$ NAE has an NP-complete CSP [4]. □

To conclude this brief subsection we notice that the sandwich technique can be slightly improved as follows.

**Lemma 17.** *Let $\mathbb{S}$ be a (possibly infinite) structure. If there is a positive integer $l$ such that $\text{CSP}_{>l}(\mathbb{S})$ is in P, then PCSP($\mathbb{A}, \mathbb{B}$) is in P for any structures $\mathbb{A} \to \mathbb{S} \to \mathbb{B}$.*

*Proof.* On input $\mathbb{C}$ to PCSP($\mathbb{A}, \mathbb{B}$) we construct $\mathbb{C}'$ of girth larger than $l$ such that $\mathbb{C}' \to \mathbb{A}$ if and only if $\mathbb{C} \to \mathbb{A}$, and $\mathbb{C}' \to \mathbb{B}$ if and only if $\mathbb{C} \to \mathbb{B}$ (via Theorem 6). Then, we test if $\mathbb{C}' \in \text{CSP}(\mathbb{S})$: if yes, then $\mathbb{C}' \to \mathbb{B}$, so $\mathbb{C} \to \mathbb{B}$, and (given the promise) we conclude that $\mathbb{C} \to \mathbb{A}$; if no, then $\mathbb{C}' \not\to \mathbb{A}$, so $\mathbb{C} \not\to \mathbb{A}$, and (given the promise) $\mathbb{C} \not\to \mathbb{B}$. □

Recall that, contrary to finite structures, CSP($\mathbb{S}$) and $\text{CSP}_{>l}(\mathbb{S})$ are not necessarily polynomial-time equivalent for infinite structures $\mathbb{S}$. As mentioned above, it was conjectured in [15] that if PCSP($\mathbb{A}, \mathbb{B}$) is polynomial-time solvable, then there is a structure $\mathbb{S}$ such that CSP($\mathbb{S}$) is in P and $\mathbb{A} \to \mathbb{S} \to \mathbb{B}$. In light of this conjecture and Lemma 17, it makes sense to ask the following.

**Question 18.** *Let $\mathbb{A}$ and $\mathbb{B}$ be finite structures, and $\mathbb{S}$ be such that $\mathbb{A} \to \mathbb{S} \to \mathbb{B}$. If $\text{CSP}_{>l}(\mathbb{S})$ is polynomial-time solvable for some positive integer $l$, does there exist a structure $\mathbb{S}'$ such that $\mathbb{A} \to \mathbb{S}' \to \mathbb{B}$ and CSP($\mathbb{S}'$) is in P?*

---

[1]Here, 1-IN-3 and NAE are the structures encoding positive 1-in-3 SAT and positive not-all-equal 3-SAT, respectively, i.e., 1-IN-3 = $(\{0,1\}, \{(1,0,0), (0,1,0), (0,0,1)\})$ and NAE = $(\{0,1\}, \{0,1\}^3 \setminus \{(0,0,0), (1,1,1)\})$

## 4.2 Graphs

In the scope of graph promise constraint satisfaction problems, it was conjectured in [14, Conjecture 1.2] that $\mathrm{PCSP}(C_{2n+1}, K_k)$ is NP-hard for every pair of positive integers $n, k$ with $k \geq 3$. It was observed in [9] that this implies that if $\mathbb{G}$ is a non-bipartite (possibly infinite) graph with finite chromatic number, then $\mathrm{CSP}(\mathbb{G})$ is NP-hard, so the authors asked whether this statement is true. Another simple application of Theorem 13 is that the previous statement is true for GMSNP.

**Theorem 19.** *Let $\mathbb{G}$ be a non-bipartite graph with finite chromatic number. If $\mathrm{CSP}(\mathbb{G})$ is expressible in GMSNP, then $\mathrm{CSP}(\mathbb{G})$ is NP-complete.*

*Proof.* By Theorem 12, there is a finite graph $\mathbb{C}$ and a positive integer $l$ such that $\mathrm{CSP}_{>l}(\mathbb{G}) = \mathrm{CSP}_{>l}(\mathbb{C})$, and by Theorem 13 $\mathrm{CSP}(\mathbb{C})$ reduces in polynomial time to $\mathrm{CSP}(\mathbb{S})$. To conclude the claim it suffices to show that $\mathbb{C}$ is a loopless non-bipartite graph. By Theorem 9, $\mathbb{C}$ is a minimal finite factor of $\mathbb{G}$, thus $\mathbb{G} \to \mathbb{C}$, and for every finite graph $\mathbb{H}$ there is a homomorphism $\mathbb{G} \to \mathbb{H}$ if and only if $\mathbb{C} \to \mathbb{H}$. Since $\mathbb{G}$ is non-bipartite, $\mathbb{C}$ is non-bipartite, and since $\mathbb{G}$ maps to a finite complete graph, then $\mathbb{C}$ is loopless. The claim now follows. □

Actually, this statement has a slightly stronger form which also follows from Theorem 13. A *smooth digraph* if a digraph with no sources nor sinks. For simplicity we say that a digraph is *hard* if $\mathrm{CSP}(\mathbb{D})$ is NP-hard. It is known that a core smooth digraph is hard whenever it is not a disjoint union of directed cycles [12].

**Corollary 20.** *Let $\mathbb{D}$ be a digraph with finite chromatic number such that $\mathbb{D}$ contains a hard smooth digraph. If $\mathrm{CSP}(\mathbb{D})$ is in GMSNP, then $\mathrm{CSP}(\mathbb{D})$ is NP-complete.*

*Proof.* Let $\mathbb{C}$ be the minimal finite factor of $\mathbb{D}$, and let $\mathbb{A}$ be a hard smooth subdigraph of $\mathbb{D}$. Then $\mathbb{A} \to \mathbb{C}$, and thus if follows that $\mathrm{CSP}(\mathbb{C})$ is NP-complete (see, e.g., [1], or [22, Theorem 5.2]). Therefore, $\mathrm{CSP}(\mathbb{D})$ is NP-complete. □

Due to the equivalent expressive power of GMSNP and forbidden pattern problems (Theorem 5), Theorem 19 can be stated combinatorially as follows.

**Corollary 21.** *Let $\mathcal{F}$ be a finite set of edge and vertex coloured graphs for which there is a positive integer $k$ such that every graph $\mathbb{G}$ that admits an $\mathcal{F}$-free colouring has chromatic number at most $k$. Then, one of the following statements hold:*

- *some odd cycle admits a colouring in $\mathrm{Forb}(\mathcal{F})$, and in this case $\mathrm{FPP}(\mathcal{F})$ is NP-complete, or*

- *all graphs that admit a colouring in $\mathrm{Forb}(\mathcal{F})$ are bipartite, and in this case $\mathrm{FPP}(\mathcal{F})$ is in P.*

Conjecture 1 asserts that any non-bipartite graph $\mathbb{G}$ as in Theorem 19 primitively positively constructs $K_3$, and thus it pp constructs its minimal finite factor $\mathbb{C}$. We ask whether this is the case for the whole scope of GMSNP (which trivially has a positive answer for finite structures $\mathbb{A}$ since $\mathbb{A}$ is its own minimal finite factor).

**Question 22.** *Does every structure $\mathbb{S}$ such that $\mathrm{CSP}(\mathbb{S})$ is in GMSNP primitively positively constructs its minimal finite factor $\mathbb{C}$?*

In general, suppose that $\mathrm{CSP}(\mathbb{S})$ is finite-domain up to high girth, and $\mathbb{C}$ a finite representative of high girth instances of $\mathrm{CSP}(\mathbb{S})$; equivalently, $\mathbb{C}$ is the minimal finite factor of $\mathbb{S}$ (Corollary 10). In this case, if $\mathrm{CSP}(\mathbb{C})$ is NP-complete, then $\mathrm{CSP}(\mathbb{S})$ is NP-complete (Remark 8), and so, the tractability conjecture [11, Conjecutre 1.2] implies a positive answer to the following question.

**Question 23.** *Let $\mathbb{S}$ be a reduct of a finitely bounded homogeneous structure such that $\mathrm{CSP}(\mathbb{S})$ is finite-domain up to high girth, and let $\mathbb{C}$ be its minimal finite factor. If $\mathrm{CSP}(\mathbb{C})$ is NP-complete, does $\mathbb{S}$ primitively positively constructs $\mathbb{C}$?*

# Acknowledgements

The author gratefully acknowledges Manuel Bodirsky, Jakub Rydval and Alexey Barsukov for valuable comments and suggestions improving the present note. In particular, to the last two for pointing out the logically equivalent (to MMSNP$_2$), but syntactically larger logic GMSNP.